\pdfoutput=1
\documentclass[11pt]{article}

\usepackage[final]{coling}

\usepackage{times}
\usepackage{latexsym}
\usepackage[T1]{fontenc}
\usepackage[utf8]{inputenc}

\usepackage{listings}
\usepackage{geometry}
\usepackage{microtype}

\usepackage{inconsolata}

\usepackage{graphicx}
\usepackage{amsmath}
\usepackage{multirow}
\usepackage{array}
\usepackage{booktabs}
\usepackage{colortbl}
\usepackage{rotating}
\usepackage{pgfplots}
\pgfplotsset{compat=1.18}

\title{OKG: On-the-Fly Keyword Generation in Sponsored Search Advertising}

\author{Zhao Wang , Briti Gangopadhyay , Mengjie Zhao , Shingo Takamatsu \\
        Sony Group Corporation\\ Tokyo, Japan}

\begin{document}
\maketitle
\begin{abstract}
Current keyword decision-making in sponsored search advertising relies on large, static datasets, limiting the ability to automatically set up keywords and adapt to real-time KPI metrics and product updates that are essential for effective advertising. In this paper, we propose On-the-fly Keyword Generation (OKG), an LLM agent-based method that dynamically monitors KPI changes and adapts keyword generation in real time, aligning with strategies recommended by advertising platforms. Additionally, we introduce the first publicly accessible dataset containing real keyword data along with its KPIs across diverse domains, providing a valuable resource for future research. Experimental results show that OKG significantly improves keyword adaptability and responsiveness compared to traditional methods. The code for OKG and the dataset are available at \url{https://github.com/sony/okg}.
\end{abstract}

\section{Introduction}

In Sponsored Search Advertising (SSA) \citep{fain2006sponsored,hillard2010improving}, advertisers bid on keywords that potential customers use in search engine queries when looking for products or services \citep{google_ads}. The highest bids and most relevant ads typically secure the best placements, appearing alongside or above search results. This approach targets users at the moment they express interest, increasing the likelihood of them visiting the advertiser’s website and making a purchase \citep{lee2018rare}.

This is where keyword decision in SSA becomes crucial \citep{google_ads}. By carefully selecting or generating relevant keywords, advertisers can ensure their ads reach users who are most likely to be interested in their offerings. Effective keyword decision not only boosts the ad’s visibility\footnote{\url{https://support.google.com/google-ads/answer/2453981?hl=en}} but also enhances its relevance\footnote{\url{https://support.google.com/google-ads/answer/6167118?hl=en}}, leading to better engagement and higher conversion rates. 

% Additionally, the size of the keyword list is critical—if the list is too small, the ad budget may not be fully utilized, and human operators must put significant effort into manually expanding the list to ensure sufficient coverage. In essence, keyword decision is the cornerstone of a successful SSA strategy.

Conventionally, keyword decisions for SSA have relied heavily on deep generation-based methods. For instance, \cite{lee2018rare} utilized a conditional GAN \cite{mirza2014conditional} to expand queries into bid keywords, while \cite{lian2019end} employed a seq2seq model \cite{sutskever2014sequence} to generate ad keywords from queries. Recently, significant advancements in LLMs \cite{gpt4,reid2024gemini} in knowledge-intensive tasks have sparked new ideas not only in keyword decision but also in other related fields such as information retrieval. \cite{ziems2023large} used GPT-3 to directly map queries to relevant document identifiers, and \cite{wang2024one} generated keywords by prompt tuning and a tree-based constrained beam search.

\begin{figure}[t]
  \centering
  \hspace*{-0.5cm} % Adjust this value as needed to move the figure left
  \includegraphics[trim={0 37cm 0 0},clip,width=1.1\columnwidth]{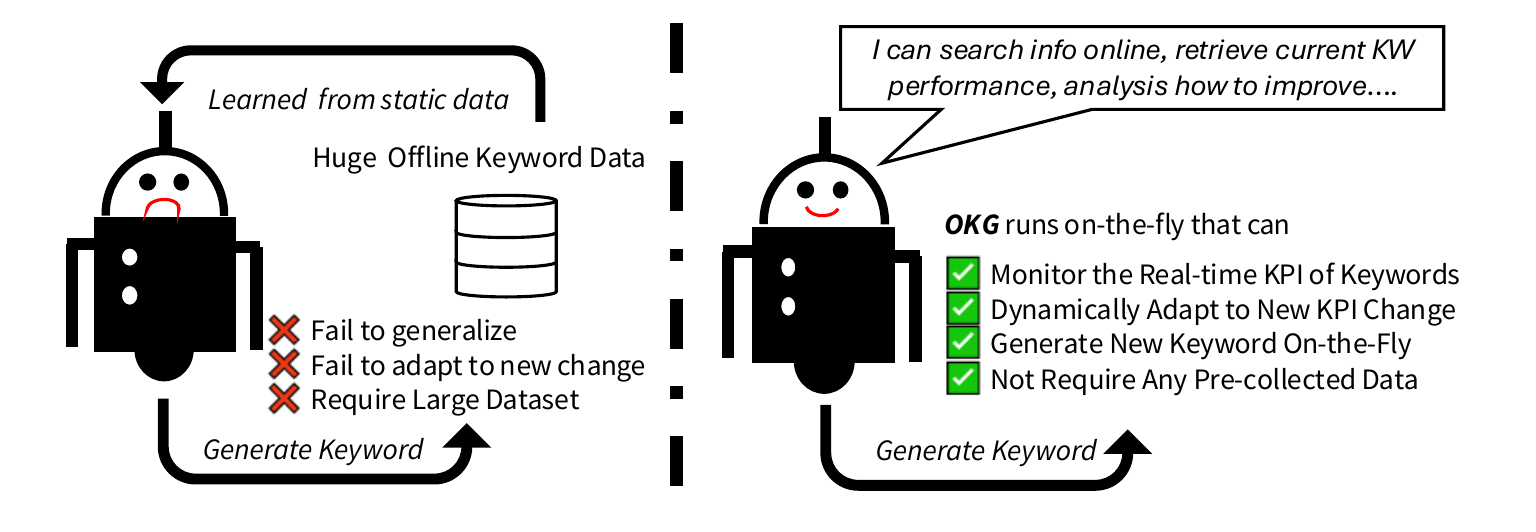}
  \caption{This visual contrasts the traditional keyword generation strategy with our OKG Agent, demonstrating the motivation behind our work.}
  \label{fig:teaser}
  \vspace{-15pt} 
\end{figure}

While both deep generation-based methods and LLM-based approaches have significantly advanced keyword generation, they come with notable drawbacks. Firstly, these methods depend on extensive keyword datasets, making them inaccessible to most advertisers who lack such data, especially given the absence of public datasets. Secondly, they fail to address the need for an adaptive, performance-driven approach in the rapidly evolving landscape of search advertising. Since both types of methods rely solely on offline data, they are inherently limited in their ability to monitor and adapt to real-time performance metrics, such as keyword clicks. This lack of real-time feedback creates inefficiencies, as models cannot adjust to performance metrics like clicks and conversions, or to rapidly changing product information. Platforms like Google\footnote{\url{https://support.google.com/google-ads/answer/1722084?hl=en}} and ad agencies emphasize the importance of continuously monitoring keyword performances\footnote{\url{https://agencyanalytics.com/blog/google-ads-metrics}} and responding to new data, such as real-time trends in user search habits, product updates, or promotions (e.g., new discounts) \citep{romer2010real}. Without this real-time adaptability, models may generate keywords that seem relevant but fail to capture current market conditions, leading to wasted ad spend and a lower return on investment.

In this paper, as shown in Fig \ref{fig:teaser}, we propose OKG, an LLM agent-based approach to SSA keyword generation that addresses the limitations of previous methods. Unlike these approaches, OKG continuously learns and evolves by observing the performance of generated keywords in live campaigns, enabling it to dynamically identify trends and optimize keyword selection. The original contributions of OKG are summarized as follows:

\begin{itemize}
    \item OKG leverages real-time information for advertising production, monitors keyword performance, and adapts automatically to changes. This capability allows the agent to judiciously expand the keyword list based on live performance data, ensuring that the keyword strategy evolves with market conditions and campaign insights.
    
    \item We propose an adaptive keyword generation method within OKG that strategically expands keywords in two dimensions: \textbf{deeper} and \textbf{wider}. The \textbf{deeper} expansion extends existing keyword categories to increase specificity and relevance, while the \textbf{wider} expansion explores new categories to capture diverse user interests and enhance campaign reach. This dual approach diversifies the keyword set while maintaining relevance, dynamically adapting to the evolving advertising landscape.
    
    \item We present a publicly accessible dataset that includes real-world Japanese keyword data with its KPIs across various domains, such as financial services, electronic devices, online shops, and AI services. This dataset is the first of its kind to be openly available, providing a valuable resource for training and evaluation in future research in SSA keyword generation.
\end{itemize}

\section{Related Works}
\label{related-work}

This section delves into the existing methodologies in SSA keyword generation, critically examining their inherent limitations and the specific challenges they fail to overcome.

\subsection{Direct Keyword Generation Using Generative Methods}
\label{subsec:generative-methods}

This section reviews two key studies that demonstrate how generative methods can directly generate keywords for sponsored search ads, showing how neural models can improve keyword generation.

\paragraph{Using GANs for Keyword Generation}
The first study by \cite{lee2018rare} uses Generative Adversarial Networks (GANs) to generate bid keywords from user queries, focusing on rare queries where traditional methods struggle. They use a sequence-to-sequence model as the generator to produce keywords based on queries, while a recurrent neural network acts as the discriminator to refine the keywords through an adversarial process.

\paragraph{NMT for Constrained Keyword Retrieval}
The second study \cite{lian2019end} applies Neural Machine Translation (NMT) to directly generate keywords from user queries in a search engine context. This end-to-end approach skips traditional steps like query rewriting. They use a Trie-based pruning technique during beam search to ensure that only valid keywords are generated, addressing the need to stay within a specific set of keywords.

\subsection{Advancements in Keyword Generation Using Large Language Models}
\label{subsec:llm-methods}

This section highlights two recent studies using Large Language Models (LLMs) for document and keyword retrieval, showing how LLMs can transform search tasks.

\paragraph{LLM for Document Retrieval}
The first study \cite{ziems2023large} overcomes the limitations of dual-encoder retrievers by using an LLM to directly generate URLs for document retrieval. Instead of encoding questions and documents separately, the LLM generates URLs by deeply interacting with user queries. By using a few {Query-URL} examples, it successfully retrieves relevant documents, with nearly 90\% accuracy in answering open-domain questions.

\paragraph{LLM for Keyword Generation in Sponsored Search}
\cite{wang2024one} presents an LLM-based keyword generation method (LKG) that treats keyword matching as an end-to-end task. Unlike traditional methods that follow a retrieve-judge-rank process, LKG uses multi-match prompt tuning, feedback tuning, and a prefix tree for constrained beam search to generate more accurate keywords.

\subsection{Limitations of Current Generative and LLM-Based Approaches}
\label{subsec:limitations}

Despite the advances in using generative and LLM-based methods for keyword generation, there are still key limitations that impact their effectiveness in dynamic search advertising.

\paragraph{Dependence on Large Datasets}
These methods often rely on access to large, proprietary query-keyword datasets, which are not available to most advertisers. Without these extensive data resources, smaller advertisers are at a disadvantage, as there are no comprehensive public datasets available.

\paragraph{Limited Real-Time Adaptability}
Most current approaches use offline data, making it hard for them to adapt to the constantly changing search advertising landscape. This lack of real-time updates means they can’t adjust quickly to changes in keyword clicks, conversions, user search behaviors, or market trends. As a result, they may generate keywords that seem relevant but don’t fit current conditions, leading to wasted ad spend and poor performance.

\paragraph{Lack of Continuous Monitoring}
Successful keyword strategies require ongoing monitoring and updates based on new data. Without this flexibility, even the most advanced models may fail to deliver optimal results in the rapidly changing world of digital advertising.

These limitations highlight the need for new methods that combine powerful modeling techniques with the ability to respond quickly to real-time data and market shifts.

\section{Problem Setting}

The task of OKG is to dynamically generate a fixed number of keywords for each time step $t$ over a time horizon $T$, where $T$ represents the total number of time steps for campaign delivery. Let $\mathcal{K}$ denote the cumulative set of all keywords generated by the end of $T$, and let $\mathcal{K}_t \subseteq \mathcal{K}$ be the specific set of keywords generated for time step $t$. Then, we have:
\[
\mathcal{K} = \bigcup_{t=1}^T \mathcal{K}_t
\]

For each time step $t$, the keyword generation process is driven by three key factors:

{
\setlength{\parindent}{0pt} % Removes indentation locally

\textbf{Information Sources ($\mathcal{S}_t$)}: Real-time data reflecting trends, product attributes, and market conditions that may change daily.

\textbf{Current Keyword Set ($\mathbf{k_t}$)}: The set of keywords generated and used during time step $t$.

\textbf{Observed KPI ($P_t$)}: The performance of the keyword set $\mathbf{k_t}$, measured by KPIs (e.g., clicks, conversions) as observed from the ad platform.
}

The keyword set for the next time step, $t+1$, is determined by OKG, denoted as $g(\mathcal{S}_t, \mathbf{k_t}, P_t)$, which considers the real-time information $\mathcal{S}_t$, the current keyword set $\mathbf{k_t}$, and its observed performance $P_t$ from time step $t$. Formally, the process is described as:
\[
\mathbf{k_{t+1}} = g(\mathcal{S}_t, \mathbf{k_t}, P_t)
\]

OKG dynamically adapts the keywords for time step $t+1$ by analyzing real-time data and adjusting based on the previous time step's performance.

The primary goal of OKG-based SSA task is to maximize the total KPI performance over the time horizon $T$, while ensuring that the number of generated keywords per time step remains fixed to optimize budget usage. The objective function is formulated as:
\[
\max_{\substack{\mathbf{k_1}, \mathbf{k_2}, \dots, \mathbf{k_T} \\ |\mathcal{K}_t| = n \, \forall t}} \sum_{t=1}^{T} P_t
\]
where $|\mathcal{K}_t| = n$ specifies that the size of the keyword set generated at each time step $t$ is fixed to $n$ keywords, which helps control the exploration of new keywords within the advertiser’s budget. Typically, advertisers operate under a fixed daily or monthly budget, so it is crucial to manage how many new keywords are explored to avoid overspending on untested keywords. 

\section{Methodology}

\begin{figure*}[h!]
\centering
\hspace*{-0.5cm} % Adjust this value as needed to move the figure left
\includegraphics[trim={0 35.8cm 0 0},clip,width=1.0\textwidth]{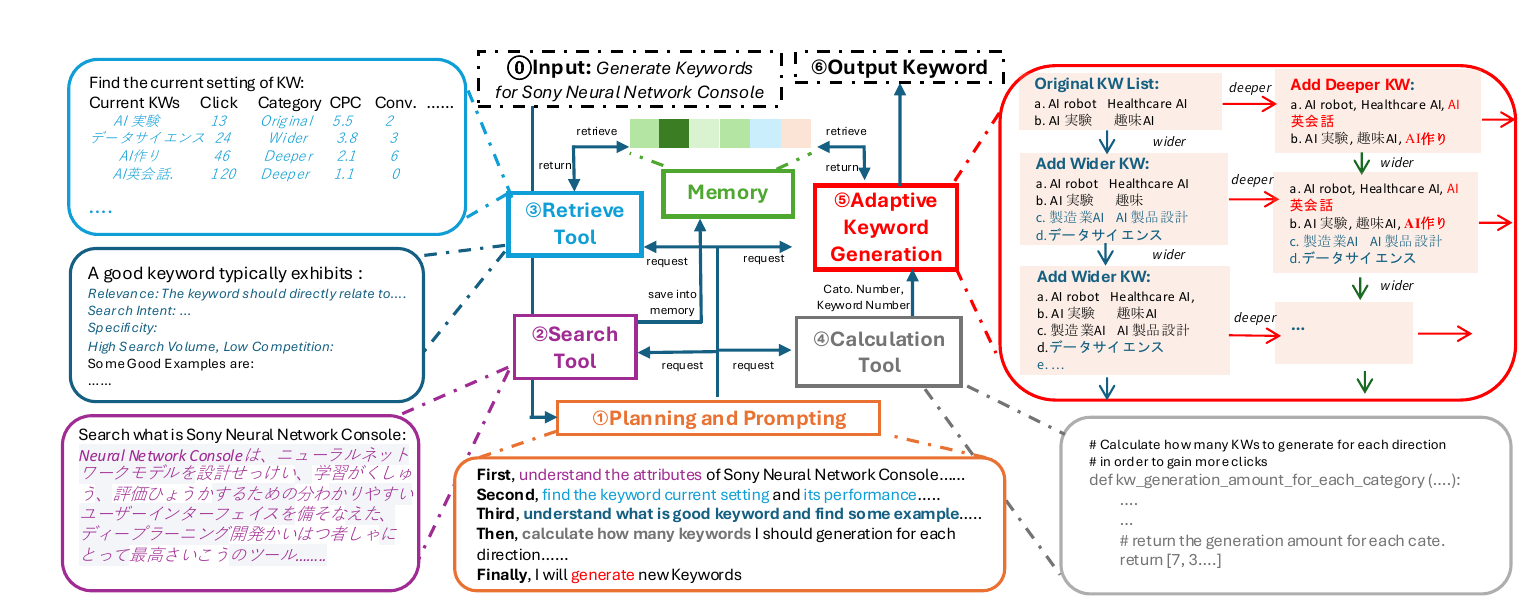}
\caption{The architecture of OKG, which fulfills the functionality of online search, real-time keyword and KPI retrieval, adaptive keyword generation, calculation and etc.}
\label{fig:okg_agent}
\vspace{-15pt} 
\end{figure*}

The architecture and workflow of OKG is illustrated in Figure~\ref{fig:okg_agent}. A detailed explanation of the key components is provided below. 

\subsection{Key Components of OKG}

\textbf{Planning and Prompting:} OKG simplifies keyword generation by eliminating the need for advertisers to gather training data or train models themselves. With just an initial prompt—``You are the expert in setting Japanese SSA keywords for \{product\}''—where the \{product\} placeholder is replaced by the specific item, OKG can automatically generate relevant keywords. This setup allows advertisers to focus on strategic elements of their campaigns, while OKG manages the technical complexities. By leveraging vast offline data, the system quickly produces high-quality keyword sets tailored to the product, reducing the cognitive load for users.

OKG also features an intelligent planning system, custom-designed to automatically plan the next steps, such as selecting the appropriate tools and identifying which KPIs ($P_t$) to monitor. Based on the initial input, OKG dynamically adjusts the keyword generation process, ensuring that the system adapts to real-time changes. An example prompt is provided in Appendix \ref{app:example}.

\paragraph{Search Tool:} The search tool \cite{serp_tool} used in OKG is responsible for gathering real-time information sources ($\mathcal{S}_t$) from the target domain. This tool retrieves data such as product attributes, current prices, discounts, and user search habits, ensuring that the generated keywords reflect the most up-to-date and accurate market conditions. For example, when generating keywords for ``Sony Neural Network Console,'' the agent retrieves live information about product specifications, pricing, and relevant search queries. This ensures that the keyword generation process is driven by real-time data ($\mathcal{S}_t$), contributing to more effective keyword strategies.

\paragraph{Retrieve and Memory Module:} OKG leverages the Google Ads API \cite{google_ads_api} to automatically gather real-time performance metrics ($P_t$), such as clicks, conversions, and other KPIs for each keyword. This real-time keyword data with its KPIs are stored in a vector-based long-term memory system \cite{faiss}, allowing for efficient tracking trends in keyword performance. The memory module organizes and stores the historical performance data ($P_{t-1}$) and new keywords generated ($\mathbf{k_t}$), ensuring that OKG can make data-driven decisions for subsequent time steps.

When OKG needs to retrieve specific information to optimize keyword strategies, it uses Retrieval-Augmented Generation (RAG) \cite{rag} to query the vector-based memory. This allows the agent to automatically access relevant historical data and real-time KPIs ($P_t$), helping it decide which keywords ($\mathbf{k_t}$) to retain, modify, or generate for the next time step. By continuously updating and retrieving information from the memory, OKG remains adaptive and responsive to changes in the advertising environment, ensuring optimal campaign effectiveness.

\subsection{Adaptive Keyword Generation with Calculation Tool}

Adaptive keyword generation is a key component of our OKG-based SSA framework, aiming to dynamically optimize keyword strategies to maximize campaign effectiveness. From \( t = 0 \), initial keywords are selected to reflect distinct product attributes, targeting various potential customer segments and adapting to market dynamics over time.

The keyword generation process is driven by two primary strategies:

\textbf{Wider Direction} (\( W_t \)): Exploring and expanding the scope by introducing new categories of keywords to capture potential new users and customer segments, \( |W_t| \) represents the number of new categories explored at time step \( t \).

\textbf{Deeper Direction} (\( D_t \)): Exploiting and intensifying focus on existing successful keyword categories, prioritizing those that have demonstrated high KPI metrics, \( |D_t| \) denotes the number of new keywords generated in the existing categories at time step \( t \).

The distribution between \( W_t \) and \( D_t \) is adaptively managed based on real-time performance data. OKG dynamically adjusts keyword generation, balancing exploration and exploitation. The keyword set for the next time step, \( \mathbf{k_{t+1}} \), is generated as:
\[
\mathbf{k_{t+1}} = g(\mathcal{S}_t, \mathbf{k_t}, P_t) = W_t \cup D_t
\]

Given the fixed size \( |\mathcal{K}_t| = n \), the distribution between \( W_t \) (new categories) and \( D_t \) (new keywords in existing categories) is determined based on the accumulated KPI from the previous time step \( P_{t-1} \). The proportion of keywords allocated to each direction is:
\[
p^W_t = \frac{P^W_{t-1}}{P_{t-1}}, \quad p^D_t = \frac{P^D_{t-1}}{P_{t-1}}
\]
\[
|W_t| = \lfloor p^W_t \cdot |\mathcal{K}_t| \rfloor, \quad |D_t| = |\mathcal{K}_t| - |W_t|
\]

This proportional allocation ensures \( |W_t| \) and \( |D_t| \) are dynamically adjusted, while maintaining the fixed total \( |\mathcal{K}_t| = n \). OKG optimizes the balance between exploring new keywords and focusing on high-performing categories, thus aligning keyword sets with emerging trends and proven preferences while controlling budget usage.

\section{Experiments}

\paragraph{Dataset.}
Considering that there are no suitable public benchmarks for training and evaluating keyword generation, we collected and sampled our real dataset from the Google Ad system over a period of six months. The dataset includes real advertisement deliveries for 10 Sony products and IT services: Sony electronic devices like cameras and TVs, Sony financial services including Sony Bank mortgages and health insurance, and Sony AI platforms such as the Sony Neural Network Console and Prediction One. The dataset contains not only the actual delivered keywords but also the performance of each keyword, including search volume, clicks, competitor score, and cost-per-click. The dataset is available at \url{https://github.com/sony/okg} 

\paragraph{Implementation Details.}
We deployed GPT-4 \citep{gpt4} as the LLM backbone for OKG, with the temperature set to 0.1. The final keywords are generated over a time horizon of $T = 3$. At each time step $t$, keywords are adaptively generated by allowing OKG to automatically observe real-time source information and feedback from KPI performance. We chose $T = 3$ for two main reasons: (1) A typical keyword list for one product is capped at around 100 keywords, and three iterations are sufficient to reach this limit while demonstrating the effectiveness of OKG compared to baselines; and (2) the execution time for three iterations is approximately two hours due to the complexity of OKG. As the number of iterations increases, the execution time doubles with each turn, since the keyword list expands with every iteration. OKG is implemented using the Langchain library \citep{langchain}. All experiments were conducted on a single machine with one NVIDIA V100 GPU and a 24-core Intel Xeon Gold-6271 processor clocked at 2.60 GHz.

\paragraph{Baselines.}
We consider the following three types of baselines:

{
\setlength{\parindent}{0pt} % Removes indentation locally

\textbf{LLM-based Baselines}, including GPT-4 \citep{gpt4} and Gemini-1.5-Pro \citep{reid2024gemini}, which are proven to be among the most powerful LLM models \citep{2024olympicarenamedalranksintelligent}.

\textbf{Japanese Keyword Extractor-based Baselines}, including Choi \citep{boolway} and RAKE \citep{RAKE}.

\textbf{Existing Commercial Application}, including Google Keyword Planner \citep{google_keyword_planner}, as baselines for our comparison.

}

{
\setlength{\parindent}{0pt} % Removes indentation locally
\subsection{Comparison on Keyword Performance}

We evaluate OKG on real keyword KPIs using the following four metrics:

\textbf{Click:} A higher click count typically indicates greater user engagement, making it a crucial indicator of keyword success.

\textbf{Search Volume:} This metric assesses keyword popularity and demand.

\textbf{Cost Per Click (CPC):} The average cost paid for each click on a keyword. CPC is vital for gauging the financial efficiency of keyword strategies, reflecting the cost-effectiveness of each click.

\textbf{Competitor Score:} A measure of market competitiveness for a keyword. It considers the number of advertisers bidding on the keyword and the bid amounts, providing a snapshot of the competitive environment.
}

The KPIs are obtained from our public dataset. Generated and original keywords are tokenized and embedded (using pooled embeddings) from a pretrained multilingual BERT model \cite{google_multilingual_bert} to measure cosine similarity. For each generated keyword, we select the most similar keyword from offline data (highest similarity score and cosine similarity > 0.6) and use its KPIs to represent the generated keyword’s KPIs.

We do not include conversion rates or other downstream metrics like Return on Ad Spend (ROAS) in our evaluation, as these metrics are highly influenced by factors beyond keyword performance alone—such as brand reputation, the quality of landing pages, and varying ad spend strategies across industries (e.g., real estate advertisers may prioritize high spending per conversion). These external variables introduce inconsistencies, making it challenging to attribute performance purely to the effectiveness of the keywords themselves.

Table \ref{tab1:general} compares keyword performance across baseline methods. OKG consistently outperforms others in key metrics such as Clicks, CPC, and Competitor Score, proving its effectiveness in optimizing keyword performance. While OKG shows lower search volume, this should be interpreted cautiously, as higher volumes don’t always translate to better relevance or clicks. OKG’s niche, targeted keywords often better match user intent and offer higher value despite lower competition.

\begin{table}[t]
\caption{Comparison on Real Keyword Performance. Clicks, Search Volumes and CPC are normalized (with N. in column name) to overcome the impact of scale differences across different products.}
\label{tab1:general}
\centering
\scriptsize
\begin{tabular}{p{0.5cm}p{0.70cm}cccc}
\toprule
\multicolumn{2}{c}{Baselines} & \multicolumn{4}{c}{Keyword Performance} \\
\cmidrule(lr){1-2} \cmidrule(lr){3-6}
Cat. & Name & Click $\uparrow$ & \begin{tabular}[c]{@{}c@{}}Srch.\\Vol.\end{tabular} $\uparrow$ & CPC $\downarrow$ & \begin{tabular}[c]{@{}c@{}}Comp.\\Score\end{tabular} $\downarrow$ \\
& & N.\textit{(0$\sim$100)} & N.\textit{(0$\sim$100)} & N.\textit{(0$\sim$1)} & \textit{(0$\sim$100)} \\
\midrule
\multirow{3}{*}{LLM}  
                      & OKG &  \textbf{100.0} & 62.3 & \textbf{0.38} & \textbf{56} \\
                      & GPT4 & 76.2 & \textbf{100.0} & 0.63 & 78 \\
                      & Gemini1.5 & 69.1 & 57.30 & 0.62 & 83 \\
\midrule
\multirow{2}{*}{\begin{tabular}[c]{@{}l@{}}Kwd.\\Ext.\end{tabular}}  
                      & Choi & 71.8 & 65.7 & 0.76 & 79 \\
                      & RAKE & 69.8 & 55.87 & 0.87 & 80 \\
\midrule 
App. & \begin{tabular}[c]{@{}c@{}}Google\\KW Plnr.\end{tabular} & 44.2 & 43 & 1.0 & 67 \\
\bottomrule
\end{tabular}
\end{table}

\subsection{Comparison on Online Relevance}
As OKG generates keywords based on online searches and real-time information using search tools, this section evaluates the generated keyword lists to determine their effectiveness in covering the information presented in search results.

{
\setlength{\parindent}{0pt} % Removes indentation locally

To accurately measure the coverage and relevance of the generated keywords, we employ several established metrics:

\textbf{BLEU-2 \cite{papineni-etal-2002-bleu}}: to measure the overlap between the generated keywords and the online search results, providing insights into how well the keywords match actual search queries;

\textbf{ROUGE-1 \cite{lin-2004-rouge}}: to focus on recall by comparing the common n-grams between the generated keywords and the target search results, indicating the extent to which our keywords capture the necessary information;

\textbf{BERTScore \cite{zhang2019bertscore}}: to assess semantic similarity, offering a deeper understanding of how effectively the generated keywords encompass the nuances of the information presented.
}

\begin{table}[t]
\caption{Comparison on Relevance and Coverage with Source Meta-data.}
\label{tab1:source_compare}
\centering
\scriptsize
\begin{tabular}{lcccc}
\toprule
\multicolumn{2}{c}{Baselines} & Relevance & \multicolumn{2}{c}{Coverage} \\
\cmidrule(lr){1-2} \cmidrule(lr){3-3} \cmidrule(lr){4-5}
Category & Name & Bert-Score $\uparrow$ & Bleu2 $\uparrow$ & Rouge1 $\uparrow$ \\
\midrule
\multirow{3}{*}{LLM}  & OKG & \textbf{0.63}  & \textbf{0.27} & \textbf{0.42}  \\
                      & GPT4 & 0.61 & 0.12 & 0.23  \\
                      & Gemini1.5 & 0.59 & 0.13 & 0.21  \\
\midrule
\multirow{2}{*}{\begin{tabular}[c]{@{}l@{}}Kwd.\\Ext.\end{tabular}}  
                      & Choi & 0.45 & 0.14 & 0.22  \\
                      & RAKE & 0.48 & 0.16 & 0.23  \\
\midrule 
App. & \begin{tabular}[c]{@{}c@{}}Google\\KW Plnr.\end{tabular} & 0.40 & 0.12 & 0.19  \\
\bottomrule
\end{tabular}
\end{table}

Table~\ref{tab1:source_compare} compares the performance of OKG with various baselines, demonstrating that OKG achieves higher relevance and coverage metrics, as measured by BERTScore, BLEU-2, and ROUGE-1, indicating the superior accuracy of OKG in generating relevant and comprehensive keywords. It is important to note that BLEU-2 and ROUGE-1 scores are relatively low across all models, including OKG, due to the inherent differences between text-to-text evaluation (for which these metrics were designed) and our text-to-keyword list evaluation.

\subsection{Comparison on Similarity with Offline Real Keywords}
{
\setlength{\parindent}{0pt} % Removes indentation locally

To evaluate the alignment between the generated keywords and real ad delivery data, we employ three key metrics:

\textbf{Jaccard similarity}: to measure the overlap between the generated keyword sets  and the real keyword sets, providing a ratio of common keywords to the union of both sets;

\textbf{Cosine similarity}: to assess the vector-based similarity between the generated keywords and real ad keywords, indicating how directionally similar the keyword sets are in the embedding space;

\textbf{BERTScore}: to evaluate the semantic similarity between the generated keywords and the real ad keywords, offering insights into how closely the meaning of the generated keywords matches the real-world data.
}

Table \ref{tab:kw_data} presents the comparison between OKG and the baselines. As shown in the table, OKG consistently outperforms the baselines across all three metrics. In particular, OKG achieves the highest BERTScore, indicating that the generated keyword lists are semantically more similar to the offline data. Similarly, OKG records superior results in both Jaccard similarity and cosine similarity, further demonstrating that our generated keywords align more closely with the real ad delivery data. These results confirm the effectiveness of OKG in generating highly relevant keywords compared to existing baselines.

\begin{table}[t]
\caption{Comparison on Similarity with Offline Real Keywords.}
\label{tab:kw_data}
\centering
\scriptsize
\begin{tabular}{lcccc}
\toprule
\multicolumn{2}{c}{Baselines} & \multicolumn{3}{c}{Offline Similarity} \\
\cmidrule(lr){1-2} \cmidrule(lr){3-5}

Category & Name & Bert-Score $\uparrow$ & Jacard $\uparrow$ & Cosine $\uparrow$ \\
\midrule
\multirow{3}{*}{LLM}  & OKG & \textbf{0.85} & \textbf{0.35} & \textbf{0.90} \\
                      & GPT4 & 0.72 & 0.30 & 0.78  \\
                      & Gemini1.5 & 0.71 & 0.28 & 0.72  \\
\midrule
\multirow{2}{*}{\begin{tabular}[c]{@{}l@{}}Kwd.\\Ext.\end{tabular}}  
                      & Choi & 0.62 & 0.22 & 0.67  \\
                      & RAKE & 0.70 & 0.25 & 0.58  \\
\midrule 
App. & \begin{tabular}[c]{@{}c@{}}Google\\KW Plnr.\end{tabular} & 0.54 & 0.20 & 0.55  \\
\bottomrule
\end{tabular}
\end{table}

\subsection{Ablation Study}

\begin{figure}[ht]
\centering
\begin{tikzpicture}

% First Y-axis for Scores (BERTScore, Jaccard, CosineSimilarity, Clicks)
\begin{axis}[
    ybar,
    symbolic x coords={BERTScore, Jaccard, Cos.Sim., Clicks},
    xtick=data,
    ymin=0, ymax=1,
    bar width=6pt,  % Further reduced bar width for narrower bars
    enlarge x limits=0.15,  % Further reduce space between bars
    axis y line*=left,
    nodes near coords,
    every node near coord/.append style={font=\tiny, scale=0.6, yshift=0.1cm},
    tick label style={font=\scriptsize},   
    legend style={at={(0.5,-0.4)}, anchor=north, legend columns=3, font=\tiny}, % Adjusted to two rows
    legend image post style={xscale=0.5},
    height=3.5cm,
    width=\columnwidth,  % Set figure width to 70% of the column
    ]

% Plot for Scores (set Clicks to zero in first axis)
\addplot[fill=pink] coordinates {(BERTScore,0.83) (Jaccard,0.34) (Cos.Sim.,0.89) (Clicks,0)};
\addplot[fill=yellow] coordinates {(BERTScore,0.81) (Jaccard,0.28) (Cos.Sim.,0.83) (Clicks,0)};
\addplot[fill=lime] coordinates {(BERTScore,0.78) (Jaccard,0.23) (Cos.Sim.,0.81) (Clicks,0)};
\addplot[fill=orange] coordinates {(BERTScore,0.72) (Jaccard,0.12) (Cos.Sim.,0.72) (Clicks,0)};
\addplot[fill=violet] coordinates {(BERTScore,0.81) (Jaccard,0.30) (Cos.Sim.,0.85) (Clicks,0)};
\legend{Full OKG, Fixed Growth OKG, Wide Growth Only, Deep Growth Only, OKG with Reflection}

\end{axis}

% Second Y-axis for Clicks
\begin{axis}[
    ybar,
    symbolic x coords={BERTScore, Jaccard, Cos.Sim., Clicks}, % Must match the first axis
    xtick=data,
    ymin=0, ymax=18000,  % Adjusted ymax to fit Clicks data
    bar width=6pt,  % Matches reduced bar width of first axis
    enlarge x limits=0.15,  % Same reduced space between bars
    axis y line*=right,
    axis x line=none,
    ytick={0, 6000, 12000, 18000},  % Adjusted for Clicks
    every node near coord/.append style={font=\tiny, scale=0.6, yshift=0.1cm},
    tick label style={font=\scriptsize},
    nodes near coords,
    height=3.5cm,
    width=\columnwidth,  % Set figure width to 70% of the column
    ]

% Plot for Clicks (set Scores to zero in second axis)
\addplot[fill=pink] coordinates {(Clicks,13998)};
\addplot[fill=yellow] coordinates { (Clicks,9803)};
\addplot[fill=lime] coordinates { (Clicks,7934)};
\addplot[fill=orange] coordinates { (Clicks,6543)};
\addplot[fill=violet] coordinates {(BERTScore,0) (Jaccard,0) (Cos.Sim.,0) (Clicks,11200)};

\end{axis}

\end{tikzpicture}
\caption{Comparison Results of Component Ablation.}
\label{fig:okg_comparison}
\end{figure}
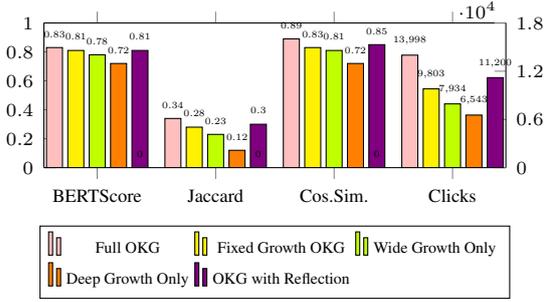

The final experiment consists of an ablation study to assess the impact of various components within the OKG framework. We performed five ablation tests on Sony TV keyword data in our dataset:

{
\setlength{\parindent}{0pt} % Removes indentation locally
\textbf{Full OKG}: The complete model with adaptive keyword generation.

\textbf{OKG with Fixed Growth}: The keyword generation process is fixed, using predefined proportions for both exploration (wider growth) and exploitation (deeper growth).

\textbf{Wide Growth Only}: Only the exploration (wider growth) mechanism is enabled.

\textbf{Deep Growth Only}: Only the exploitation (deeper growth) mechanism is enabled.

\textbf{OKG with Reflection}: Incorporates Reflexion \cite{shinn2024reflexion} feedback from previous time steps to guide future keyword generation.
}

Figure \ref{fig:okg_comparison} shows the performance across key metrics, highlighting that the Full OKG consistently outperforms the other versions. Fixing the growth directions in the OKG with Fixed Growth version results in a notable performance decline. Both Wide Growth Only and Deep Growth Only confirm that neither exploration nor exploitation alone is as effective as their combination. Interestingly, OKG with Reflexion \cite{shinn2024reflexion}, which learns from past experiences, does not yield improvements in keyword relevance, supporting our hypothesis that real-time feedback monitoring is more critical than relying on past data.

\section*{Conclusion}

We introduced OKG, a dynamic framework leveraging LLM agent to adaptively generate keywords for sponsored search advertising. Additionally, we provided the first publicly accessible dataset with real ad keyword data, offering a valuable resource for future research in keyword optimization. Experimental results and ablation studies demonstrate the effectiveness of OKG, showing significant improvements across various metrics and emphasizing the importance of each component.

\bibliography{custom}

\begin{thebibliography}{32}
\providecommand{\natexlab}[1]{#1}

\bibitem[{Achiam et~al.(2023)Achiam, Adler, Agarwal, Ahmad, Akkaya, Aleman, Almeida, Altenschmidt, Altman, Anadkat et~al.}]{gpt4}
Josh Achiam, Steven Adler, Sandhini Agarwal, Lama Ahmad, Ilge Akkaya, Florencia~Leoni Aleman, Diogo Almeida, Janko Altenschmidt, Sam Altman, Shyamal Anadkat, et~al. 2023.
\newblock Gpt-4 technical report.
\newblock \emph{arXiv preprint arXiv:2303.08774}.

\bibitem[{Choi(2024)}]{boolway}
Choi. 2024.
\newblock Keyword extractor.
\newblock \url{https://choimitena.com/Nihongo/Analyze}.
\newblock Choi.

\bibitem[{Contributors(2024)}]{langchain}
Langchain Contributors. 2024.
\newblock Langchain: Build context-aware reasoning applications with language models.
\newblock \url{https://github.com/langchain/langchain}.
\newblock Version x.x.

\bibitem[{Fain and Pedersen(2006)}]{fain2006sponsored}
Daniel~C Fain and Jan~O Pedersen. 2006.
\newblock Sponsored search: A brief history.
\newblock \emph{Bulletin-American Society For Information Science And Technology}, 32(2):12.

\bibitem[{Google(2024{\natexlab{a}})}]{google_ads}
Google. 2024{\natexlab{a}}.
\newblock \href {https://support.google.com/google-ads/answer/2472712?hl=en} {About adjusting your keyword bids}.
\newblock Accessed: 2024-09-17.

\bibitem[{Google(2024{\natexlab{b}})}]{google_multilingual_bert}
Google. 2024{\natexlab{b}}.
\newblock Bert multilingual model.
\newblock \url{https://github.com/google-research/bert/blob/master/multilingual.md}.
\newblock Accessed: 2024-09-30.

\bibitem[{{Google}(2024)}]{google_ads_api}
{Google}. 2024.
\newblock Google ads api.
\newblock \url{https://developers.google.com/google-ads/api}.
\newblock Accessed: 2024-09-25.

\bibitem[{Google(2024)}]{google_keyword_planner}
Google. 2024.
\newblock Google keyword planner.
\newblock \url{https://ads.google.com/home/tools/keyword-planner/}.
\newblock Google Ads.

\bibitem[{Hillard et~al.(2010)Hillard, Schroedl, Manavoglu, Raghavan, and Leggetter}]{hillard2010improving}
Dustin Hillard, Stefan Schroedl, Eren Manavoglu, Hema Raghavan, and Chirs Leggetter. 2010.
\newblock Improving ad relevance in sponsored search.
\newblock In \emph{Proceedings of the third ACM international conference on Web search and data mining}, pages 361--370.

\bibitem[{Huang et~al.(2024)Huang, Wang, Xia, and Liu}]{2024olympicarenamedalranksintelligent}
Zhen Huang, Zengzhi Wang, Shijie Xia, and Pengfei Liu. 2024.
\newblock \href {https://arxiv.org/abs/2406.16772} {Olympicarena medal ranks: Who is the most intelligent ai so far?}
\newblock \emph{Preprint}, arXiv:2406.16772.

\bibitem[{Johnson et~al.(2024)Johnson, Douze, and Jégou}]{faiss}
Jeff Johnson, Matthijs Douze, and Hervé Jégou. 2024.
\newblock Billion-scale similarity search with gpus.
\newblock \url{https://github.com/facebookresearch/faiss}.
\newblock Accessed: 2024-09-25.

\bibitem[{Lee et~al.(2018)Lee, Gao, and Zhang}]{lee2018rare}
Mu-Chu Lee, Bin Gao, and Ruofei Zhang. 2018.
\newblock Rare query expansion through generative adversarial networks in search advertising.
\newblock In \emph{Proceedings of the 24th acm sigkdd international conference on knowledge discovery \& data mining}, pages 500--508.

\bibitem[{Lewis et~al.(2020)Lewis, Perez, Piktus, Petroni, Karpukhin, Goyal, Küttler, Lewis, Yih, Rocktäschel, Riedel, and Kiela}]{rag}
Patrick Lewis, Ethan Perez, Aleksandra Piktus, Fabio Petroni, Vladimir Karpukhin, Naman Goyal, Heinrich Küttler, Mike Lewis, Wen-tau Yih, Tim Rocktäschel, Sebastian Riedel, and Douwe Kiela. 2020.
\newblock \href {https://arxiv.org/abs/2005.11401} {Retrieval-augmented generation for knowledge-intensive nlp tasks}.
\newblock In \emph{Advances in Neural Information Processing Systems}.

\bibitem[{Lian et~al.(2019)Lian, Chen, Hu, Zhang, Yan, Tong, Han, Guan, Li, Cao et~al.}]{lian2019end}
Yijiang Lian, Zhijie Chen, Jinlong Hu, Kefeng Zhang, Chunwei Yan, Muchenxuan Tong, Wenying Han, Hanju Guan, Ying Li, Ying Cao, et~al. 2019.
\newblock An end-to-end generative retrieval method for sponsored search engine--decoding efficiently into a closed target domain.
\newblock \emph{arXiv preprint arXiv:1902.00592}.

\bibitem[{Lin(2004)}]{lin-2004-rouge}
Chin-Yew Lin. 2004.
\newblock \href {https://aclanthology.org/W04-1013} {{ROUGE}: A package for automatic evaluation of summaries}.
\newblock In \emph{Text Summarization Branches Out}, pages 74--81, Barcelona, Spain. Association for Computational Linguistics.

\bibitem[{Mirza and Osindero(2014)}]{mirza2014conditional}
Mehdi Mirza and Simon Osindero. 2014.
\newblock Conditional generative adversarial nets.
\newblock \emph{arXiv preprint arXiv:1411.1784}.

\bibitem[{Nie et~al.(2019)Nie, Yang, and Zeng}]{nie2019keyword}
Han Nie, Yanwu Yang, and Daniel Zeng. 2019.
\newblock Keyword generation for sponsored search advertising: Balancing coverage and relevance.
\newblock \emph{IEEE intelligent systems}, 34(5):14--24.

\bibitem[{Papineni et~al.(2002)Papineni, Roukos, Ward, and Zhu}]{papineni-etal-2002-bleu}
Kishore Papineni, Salim Roukos, Todd Ward, and Wei-Jing Zhu. 2002.
\newblock \href {https://doi.org/10.3115/1073083.1073135} {{B}leu: a method for automatic evaluation of machine translation}.
\newblock In \emph{Proceedings of the 40th Annual Meeting of the Association for Computational Linguistics}, pages 311--318, Philadelphia, Pennsylvania, USA. Association for Computational Linguistics.

\bibitem[{Reid et~al.(2024)Reid, Savinov, Teplyashin, Lepikhin, Lillicrap, Alayrac, Soricut, Lazaridou, Firat, Schrittwieser et~al.}]{reid2024gemini}
Machel Reid, Nikolay Savinov, Denis Teplyashin, Dmitry Lepikhin, Timothy Lillicrap, Jean-baptiste Alayrac, Radu Soricut, Angeliki Lazaridou, Orhan Firat, Julian Schrittwieser, et~al. 2024.
\newblock Gemini 1.5: Unlocking multimodal understanding across millions of tokens of context.
\newblock \emph{arXiv preprint arXiv:2403.05530}.

\bibitem[{R{\"o}mer et~al.(2010)R{\"o}mer, Ostermaier, Mattern, Fahrmair, and Kellerer}]{romer2010real}
Kay R{\"o}mer, Benedikt Ostermaier, Friedemann Mattern, Michael Fahrmair, and Wolfgang Kellerer. 2010.
\newblock Real-time search for real-world entities: A survey.
\newblock \emph{Proceedings of the IEEE}, 98(11):1887--1902.

\bibitem[{Rose et~al.(2010)Rose, Engel, Cramer, and Cowley}]{RAKE}
Stuart Rose, Dave Engel, Nick Cramer, and Wendy Cowley. 2010.
\newblock Automatic keyword extraction from individual documents.
\newblock \emph{Text mining: applications and theory}, pages 1--20.

\bibitem[{Scholz et~al.(2019)Scholz, Brenner, and Hinz}]{scholz2019akegis}
Michael Scholz, Christoph Brenner, and Oliver Hinz. 2019.
\newblock Akegis: automatic keyword generation for sponsored search advertising in online retailing.
\newblock \emph{Decision Support Systems}, 119:96--106.

\bibitem[{{Serp}(2024)}]{serp_tool}
{Serp}. 2024.
\newblock Serp: Real-time search engine data for seo and marketing.
\newblock \url{https://serpapi.com/}.
\newblock Accessed: 2024-09-25.

\bibitem[{Shinn et~al.(2024)Shinn, Cassano, Gopinath, Narasimhan, and Yao}]{shinn2024reflexion}
Noah Shinn, Federico Cassano, Ashwin Gopinath, Karthik Narasimhan, and Shunyu Yao. 2024.
\newblock Reflexion: Language agents with verbal reinforcement learning.
\newblock \emph{Advances in Neural Information Processing Systems}, 36.

\bibitem[{Sutskever(2014)}]{sutskever2014sequence}
I~Sutskever. 2014.
\newblock Sequence to sequence learning with neural networks.
\newblock \emph{arXiv preprint arXiv:1409.3215}.

\bibitem[{Wang et~al.(2024)Wang, Sha, Lin, Feng, Zhu, Wang, Jiao, Huang, Ye, He et~al.}]{wang2024one}
Yang Wang, Zheyi Sha, Kunhai Lin, Chaobing Feng, Kunhong Zhu, Lipeng Wang, Xuewu Jiao, Fei Huang, Chao Ye, Dengwu He, et~al. 2024.
\newblock One-step reach: Llm-based keyword generation for sponsored search advertising.
\newblock In \emph{Companion Proceedings of the ACM on Web Conference 2024}, pages 1604--1608.

\bibitem[{Yang and Li(2023)}]{yang2023keyword}
Yanwu Yang and Huiran Li. 2023.
\newblock Keyword decisions in sponsored search advertising: A literature review and research agenda.
\newblock \emph{Information Processing \& Management}, 60(1):103142.

\bibitem[{Zhang and Qiao(2018)}]{zhang2018novel}
Jin Zhang and Dandan Qiao. 2018.
\newblock A novel keyword suggestion method for search engine advertising.
\newblock \emph{IEEE intelligent systems}.

\bibitem[{Zhang et~al.(2023)Zhang, Zhang, and Chen}]{zhang2023semantic}
Jin Zhang, Jilong Zhang, and Guoqing Chen. 2023.
\newblock A semantic transfer approach to keyword suggestion for search engine advertising.
\newblock \emph{Electronic Commerce Research}, pages 1--27.

\bibitem[{Zhang et~al.(2019)Zhang, Kishore, Wu, Weinberger, and Artzi}]{zhang2019bertscore}
Tianyi Zhang, Varsha Kishore, Felix Wu, Kilian~Q Weinberger, and Yoav Artzi. 2019.
\newblock Bertscore: Evaluating text generation with bert.
\newblock \emph{arXiv preprint arXiv:1904.09675}.

\bibitem[{Zhou et~al.(2019)Zhou, Huang, Mao, Zhu, Shu, and Zhu}]{zhou2019domain}
Hao Zhou, Minlie Huang, Yishun Mao, Changlei Zhu, Peng Shu, and Xiaoyan Zhu. 2019.
\newblock Domain-constrained advertising keyword generation.
\newblock In \emph{The World Wide Web Conference}, pages 2448--2459.

\bibitem[{Ziems et~al.(2023)Ziems, Yu, Zhang, and Jiang}]{ziems2023large}
Noah Ziems, Wenhao Yu, Zhihan Zhang, and Meng Jiang. 2023.
\newblock Large language models are built-in autoregressive search engines.
\newblock \emph{arXiv preprint arXiv:2305.09612}.

\end{thebibliography}

\appendix

\clearpage
\section{An Example of OKG Generation Prompt}
\label{app:okg-detail}

\begin{figure}
    \centering
    \includegraphics[width=\columnwidth]{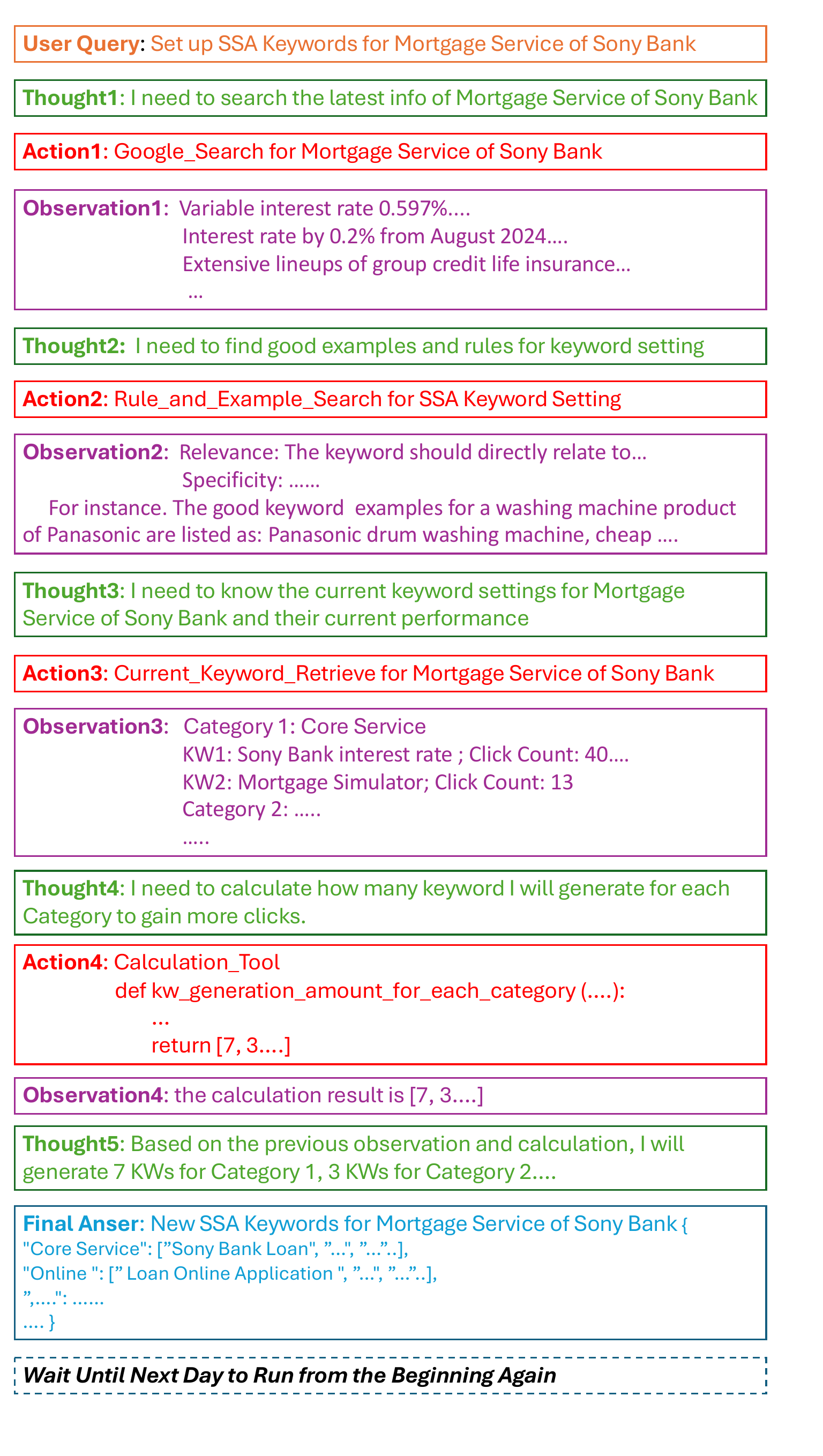}
    \caption{An intuitive example of OKG generation prompt for Sony Bank's Mortgage Service}
    \label{fig:okg_prompt_example}
\end{figure}

In this section, we provide an intuitive example illustrating how OKG generates keyword suggestions through a structured, multi-step prompt as shown in Figure~\ref{fig:okg_prompt_example}.

\subsection*{Query Understanding}
\label{subsec:query-understanding}
The process begins with a user query to set up SSA keywords for Mortgage Service of Sony Bank. OKG parses this query to understand the specific requirements---such as the product focus (mortgage services) and the target entity (Sony Bank).

\subsection*{Step 1: Gathering Current Market Data}
\begin{description}
    \item[Action:] The system performs a Google search to gather the latest relevant information about Sony Bank's mortgage services.
    \item[Observation:] It notes the current interest rates, insurance options, and other service features that are critical for keyword relevance.
\end{description}

\subsection*{Step 2: Benchmarking Against Practices}
\begin{description}
    \item[Action:] OKG queries databases and previous case studies for effective keyword strategies in similar sectors.
    \item[Observation:] It identifies key attributes like relevance and specificity, which are crucial for the effectiveness of the keywords. 
\end{description}

\subsection*{Step 3: Analyzing Current Keyword Performance}
\begin{description}
    \item[Action:] The system retrieves and analyzes performance data of existing keywords related to Sony Bank’s mortgage services.
    \item[Observation:] Keywords are categorized by categories, such as click counts and search volumes. This data helps in understanding which types of keywords are currently performing well.
\end{description}

\subsection*{Step 4: Strategic Keyword Generation}
\begin{description}
    \item[Action:] Based on the collected data and observed patterns, OKG calculates the optimal number of new keywords to generate for each category.
    \item[Observation:] The decision on the quantity of keywords is influenced by their potential to improve click-through rates and overall campaign performance.
\end{description}

\subsection*{Step 5: Generating and Implementing New Keywords}
\begin{description}
    \item[Outcome:] Utilizing the insights gained from the above steps, OKG generates a tailored list of new keywords..
\end{description}

\subsection*{Iterative Refinement Everytime span}
The process is inherently iterative, allowing for continuous refinement and optimization. OKG's ability to adapt to dynamic market conditions and shifting user preferences stands as a key differentiator in its operational efficacy.

\clearpage
\section{Related Works of SSA Keyword Dataset}
\label{app:related-work--data}

A review of recent literature \cite{yang2023keyword} reveals a dependence on diverse, predominantly private datasets for training and validating keyword generation models. In recent several years, \cite{zhang2018novel} utilized query logs collected through the Google Keyword Suggestion Tool, focusing on query keywords and query volumes for seed keywords. \cite{nie2019keyword} constructed their dataset by crawling Wikipedia, which, while extensive, was confined to the context of content generation and not specific commercial keyword use. \cite{scholz2019akegis} documented SSA campaign performances for large-scale online retailers provided by a company with significant online sales, highlighting the commercial and proprietary nature of the dataset. \cite{zhou2019domain} employed 40 million query logs from Sogou.com, with each sample consisting of a keyword and user query pair, reflecting real-world business queries but not publicly available for research. Similarly, \cite{zhang2023semantic} analyzed query logs and keyword performance through private datasets that detail interactions but are not accessible to the public. \cite{wang2024one} generated keyword by prompt tuning and a tree-based constrained beam search based on a private dataset.

These cases underline a prevalent issue in the field: a significant reliance on private or business-specific data. The absence of publicly accessible datasets not only hampers the reproducibility of research but also limits the development of SSA keyword generation models that could benefit a wider range of advertisers, particularly those without access to large-scale data repositories. This limitation in data accessibility motivates the development of our OKG, which aims to bridge these gaps by introducing a novel dataset that is both publicly accessible and robust enough to train effective SSA models.

The need for publicly accessible datasets is underscored by the growing demand for transparency and reproducibility in machine learning research, as discussed by various scholars and echoed in recent AI ethics guidelines.

\section{An Example of Generated Keyword by OKG}
\label{app:example}

In this section, we present an example of keywords generated by our proposed OKG. Starting from initial keywords, we demonstrate how new keywords are generated in each round. 

\textbf{Note:} For readability and to cater to an English-speaking audience, the original generated keywords in Japanese have been translated into English.

\subsection{Keywords Generated By OKG}

\textbf{Initialized Keywords}

\begin{lstlisting}
{
    "Illness Coverage Domains": ["Cancer Insurance"],
    "Core Service": ["Medical Insurance"],
    "Attribute": ["Cheap Medical Insurance"]
}
\end{lstlisting}

\textbf{First Round}

\begin{lstlisting}
{
    "Illness Coverage Domains": ["Sony Cancer Insurance", "Sickness Insurance", "Hospitalization Insurance"],
    "Core Service": ["Sony Insurance Medical", "Sony Insurance", "Sony Lifetime Medical Insurance"],
    "Attribute": ["Cheap Insurance", "Affordable Insurance Premiums", "Sony Insurance Premiums"],
    "Customer Service": ["Sony Insurance Contract", "Sony Insurance Claim", "Sony Insurance Contract Details"],
    "Insurance Benefits": ["Hospitalization Benefit", "Sony Insurance Hospitalization Benefit", "Surgery Benefit"],
    "Insurance Procedures": ["Sony Insurance Claim Procedure", "Sony Insurance Contract Details", "Sony Insurance Address Change"]
}
\end{lstlisting}

\textbf{Second Round}

\begin{lstlisting}
{
    "Illness Coverage Domains": ["Sony Insurance Sickness Insurance", "Sony Insurance Injury Insurance"],
    "Core Service": ["Sony Insurance", "Sony Insurance Plans"],
    "Attribute": ["Sony Insurance Safe Insurance", "Sony Insurance High-Quality Insurance"],
    "Customer Service": ["Sony Insurance Consultation", "Sony Insurance Inquiry"],
    "Insurance Benefits": ["Sony Insurance Sickness Coverage", "Sony Insurance Hospitalization Insurance"],
    "Insurance Procedures": ["Sony Insurance Application", "Sony Insurance Renewal"],
    "Payment Options": ["Sony Insurance Monthly Payment", "Sony Insurance Annual Payment", "Sony Insurance Installment Payment"],
    "Online Services": ["Sony Insurance Online Insurance", "Sony Insurance Web Insurance", "Sony Insurance Digital Insurance"]
}
\end{lstlisting}

\textbf{Third Round}

\begin{lstlisting}
{
    "Illness Coverage Domains": ["Sony Lifetime Medical Insurance", "Sony Medical Insurance Coverage"],
    "Core Service": ["Sony Insurance Medical Insurance", "Sony Insurance Products"],
    "Attribute": ["Peace of Mind Medical Insurance Sony", "Reliable Medical Insurance"],
    "Customer Service": ["Sony Insurance Contract Details Confirmation", "Sony Insurance Evaluation"],
    "Insurance Benefits": ["Sony Insurance Advanced Medical Benefits", "Sony Insurance Benefit Details"],
    "Insurance Procedures": ["Sony Insurance Procedures", "Sony Insurance Contract Procedures"],
    "Payment Options": ["Sony Insurance Premium Comparison", "Sony Insurance Pricing Plans"],
    "Online Services": ["Insurance Online Contract", "Sony Insurance Web Contract"],
    "Price Comparison": ["Sony Insurance Premiums", "Insurance Premium Comparison"],
    "Customer Review": ["Sony Insurance Review", "Sony Insurance Reputation"]
}
\end{lstlisting}

\subsection{Keywords Generated by Baseline Method (GPT-4)}

To compare the effectiveness of OKG, we also generated keywords using GPT-4 as a baseline. Below are the keywords produced by GPT-4 without iterative refinement.

\begin{lstlisting}
- Sony Health Insurance
- Affordable Health Coverage
- Sony Medical Insurance Plans
- Best Health Insurance Options
- Comprehensive Health Plans
- Family Health Insurance
- Individual Health Coverage
- Sony Insurance Services
- Health Insurance Quotes
- Compare Health Insurance
- Online Health Insurance
- Health Insurance Premiums
- Corporate Health Benefits
- Employee Health Insurance
- Medical Expense Coverage
- Health Protection Plans
- Sony Financial Services
- Private Health Insurance
- Health Insurance Enrollment
- Affordable Medical Insurance
- Health Insurance Policies
- Health Insurance for Families
- Health Insurance Providers
- Health Insurance Comparison
- Medical Insurance Benefits
- Sony Healthcare Services
- Best Medical Insurance Plans
- Health Insurance Offers
- Health Insurance Online
- Medical Insurance Quotes
- Cheap Health Insurance
- Health Insurance Deals
- Health Insurance Information
- Health Insurance Advice
- Health Insurance Guide
- Health Insurance Discounts
- Sony Insurance Quotes
- Health Insurance Options
- Medical Coverage Options
- Health Insurance Company
- Health Insurance Benefits
- Health Insurance Assistance
- Health Insurance Enrollment
- Affordable Health Insurance
- Sony Health Plans
- Health Coverage by Sony
- Medical Insurance Plans
- Health Insurance for Individuals
- Employee Health Benefits
- Sony Insurance Plans
\end{lstlisting}

\subsection{Analysis: Why OKG-Generated Keywords Are Better than GPT-4}

The comparison between OKG and GPT-4, based on the keyword examples provided in the previous subsection, highlights several important advantages of OKG over GPT-4 in generating more relevant and effective keywords:

\begin{itemize}
    \item \textbf{Contextual Relevance}:  
    OKG-generated keywords are more contextually relevant to the insurance domain and specific to Sony's insurance products. For example, keywords like \textit{"Sony Cancer Insurance"} and \textit{"Sony Insurance Premiums"} directly relate to the advertised products and services. In contrast, GPT-4 produces more generic keywords such as \textit{"Affordable Health Insurance"} and \textit{"Best Health Insurance Options"}, which lack specificity and brand alignment, making them less effective for targeted advertising.

    \item \textbf{Iterative Refinement}:  
    OKG’s iterative rounds of keyword generation lead to progressively refined keywords. For instance, in the second and third rounds, keywords like \textit{"Sony Insurance Application"} and \textit{"Sony Insurance Premium Comparison"} are introduced, offering more specific search terms based on previously generated keywords. GPT-4, on the other hand, generates a static list of keywords without refinement, lacking the depth and evolution seen in OKG’s iterative process.

    \item \textbf{Balanced Exploration and Exploitation}:  
    OKG demonstrates a balance between exploring new categories and deepening existing ones. In the first round, OKG introduces new categories such as \textit{"Insurance Benefits"} and \textit{"Payment Options"}, while in later rounds, it refines existing categories with more detailed keywords like \textit{"Sony Insurance Advanced Medical Benefits"} and \textit{"Sony Insurance Monthly Payment"}. GPT-4 does not offer this balance; its keywords are limited to broader categories, such as \textit{"Health Insurance Policies"} and \textit{"Corporate Health Benefits"}, which may not target niche user intents as effectively.

    \item \textbf{Targeted User Intent}:  
    OKG-generated keywords better align with user intent by including niche and long-tail keywords like \textit{"Sony Insurance Sickness Coverage"} and \textit{"Sony Insurance Hospitalization Insurance"}. These terms are likely to attract users specifically searching for Sony's insurance products. GPT-4, in contrast, produces more generic terms like \textit{"Health Insurance Quotes"} and \textit{"Online Health Insurance"}, which are too broad to effectively capture the precise needs of the target audience.

    \item \textbf{Brand-Specific Keywords}:  
    A key strength of OKG is its ability to consistently generate brand-specific keywords like \textit{"Sony Insurance"} in every round, which is essential for brand-driven advertising. GPT-4, however, lacks this focus on the Sony brand, producing more general health insurance terms, such as \textit{"Best Medical Insurance Plans"} and \textit{"Health Insurance Discounts"}. This brand specificity makes OKG’s output far more relevant for campaigns aimed at promoting Sony’s products.
\end{itemize}

In summary, the examples show that OKG outperforms GPT-4 by producing more contextually relevant, brand-specific, and refined keywords that evolve over time. OKG’s iterative approach and focus on balancing exploration with exploitation allow it to better capture user intent and optimize keyword performance, whereas GPT-4’s static, generic output is less suited for targeted, brand-specific advertising.

\end{document}